\documentclass{mem}
\usepackage{natbib}\usepackage{txfonts}\usepackage{balance}
\usepackage{graphicx}
\usepackage[a4paper]{hyperref}
\begin{document}

\newcommand{\Dnu}{\Delta \nu}
\newcommand{\acena}{\mbox{$\alpha$~Cen~A}}
\newcommand{\acenb}{\mbox{$\alpha$~Cen~B}}
\newcommand{\bvir}{\mbox{$\beta$~Vir}}
\newcommand{\muara}{\mbox{$\mu$~Ara}}
\newcommand{\eboo}{\mbox{$\eta$~Boo}}
\newcommand{\muHz}{\mbox{$\mu$Hz}}

\title{Observing solar-like oscillations: recent results}

\subtitle{}

\author{Timothy R. Bedding\inst{1} \and Hans Kjeldsen\inst{2}}

\offprints{Tim Bedding}

\institute{
School of Physics A28, University of Sydney, NSW 2006,
Australia.\\
\email{bedding@physics.usyd.edu.au}
\and
Department of Physics and Astronomy, University of Aarhus,
DK-8000 Aarhus C, Denmark.
\email{hans@phys.au.dk}
}

\authorrunning{Bedding \& Kjeldsen}

\titlerunning{Observations of solar-like oscillations}

\abstract{ We review recent progress in observations of ground-based
oscillations.  Excellent observations now exist for a few stars (\acena{}
and~B, \muara), while there is some controversy over others (Procyon,
\eboo).  We have reached the stage where single-site observations are of
limited value and where careful planning is needed to ensure the future of
asteroseismology.   } \maketitle{}

\section{Introduction}

Observations of solar-like oscillations are accumulating rapidly.  It is
amazing to recall that only a few years ago, and despite intense efforts by
several groups, we were still waiting for the first confirmed detection.
Oscillations have now been measured for many main-sequence and subgiant stars
using spectrographs such as CORALIE, ELODIE, HARPS, UCLES and UVES\@.  In
this review we concentrate on some of the most recent results -- for reviews
of earlier work, see \citet{B+C2003} and \citet{B+K2003}.

\section{Velocity versus Intensity}

\begin{figure*}[t!]
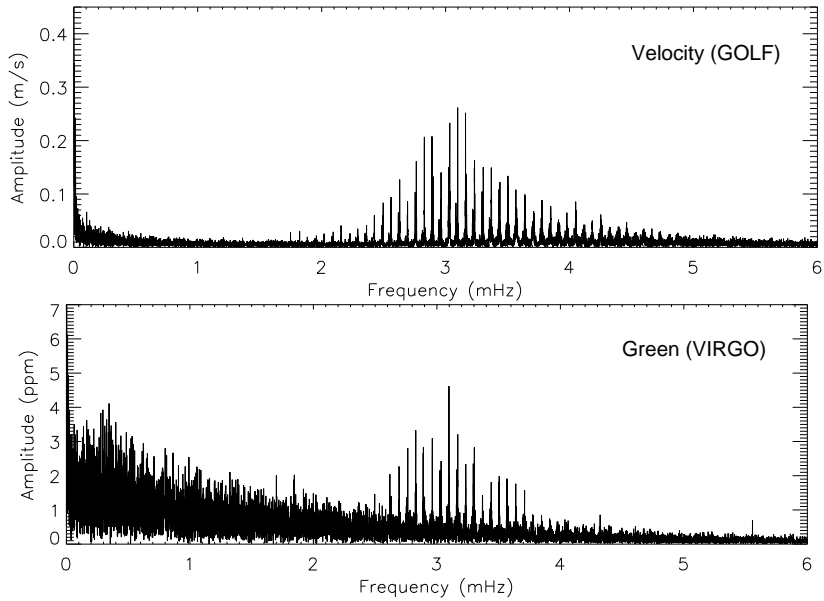

\centerline{\resizebox{0.8\hsize}{!}{\includegraphics{GOLF_1_20.epsi}}}
\centerline{\resizebox{0.78\hsize}{!}{\includegraphics{GREEN_1_20.epsi}}}
\caption{\footnotesize Amplitude spectra (square root of power) of 20 days
of full-disk observations of the Sun from the SOHO spacecraft made in
velocity (top) and intensity (bottom).  The background is due to
fluctuations from solar granulation, which is much stronger in intensity
than in velocity.  Note that the two observations were not simultaneous.  }
\label{fig.soho}
\end{figure*}

Which method is to be preferred for measuring oscillations?  It is helpful
to examine the wealth of superb oscillation data available for the Sun.
Figure~\ref{fig.soho} shows amplitude spectra from 20-d time series taken
with two different instruments on board the SOHO spacecraft: GOLF (which
measures velocity) and VIRGO (which measures intensity).  In both cases the
Sun is observed as a star, with the light being integrated over the full
disk.  Both these amplitude spectra show the regular series of peaks that
is characteristic of an oscillating sphere, but we also see a sloping
background from granulation that is much higher in intensity than in
velocity.  This background represents a fundamental noise limit and makes
velocity observations of stars potentially much more sensitive than
photometry.

Another well-known advantage of velocity measurements, although not
discernible in Fig.~\ref{fig.soho}, is that the $l=3$ modes are much
stronger.  This occurs because we measure velocities projected onto the
line of sight, which gives more sensitivity to the centre of the disk
relative to the limb and reduces the tendency for high-degree modes to
cancel out \citep{ChD+G82}.

The disadvantage of measuring velocity is that one requires a very stable
and sophisticated spectrograph.  Such spectrographs are normally mounted on
large and heavily oversubscribed telescopes, which makes it difficult to
organize multi-site campaigns.

Attempts have been made to measure solar-like oscillations in intensity,
but stellar photometry from low-altitude sites is strongly affected by
atmospheric scintillation, and the best efforts have so far fallen short of
a clear detection \citep[e.g.,][]{GBK93}.  There is hope that the high
Antarctic plateau may offer excellent conditions.  If so, the best targets
will probably be clusters of stars, and perhaps rapid rotators that are
unsuitable for Doppler measurements.  Photometry from space is unaffected
by scintillation but there are other challenges, in addition to granulation
noise, particularly from scattered light from the Sun and Earth.  An
enormous advantage of space, shared to some extent by Antarctica, is the
possibility of long continuous observing runs.

For the moment, most of the results in solar-like oscillations are coming
from ground-based Doppler measurements, with a growing input from space
telescopes such as MOST, WIRE and hopefully COROT, which is due for launch
in 2006.  In the next sections we review some of the recent results.

\section{\acena{} and B}

The clear detection of p-mode oscillations in \acena{} by \citet{B+C2002}
using the CORALIE spectrograph represented a key moment in this field.
This was followed by a dual-site campaign on this star \citep{BKB2004},
plus single-site \citep{C+B2003} and dual-site \citep{KBB2005} observations
of the B component.  The result is a set of frequencies, together with
estimates of mode lifetimes, that should keep theoreticians busy for some
time.  However, further observations with higher sensitivity are desirable
in order to measure modes with low amplitude and hence increase the number
of detected modes.

\section{Procyon}

Procyon has long been a favourite target for oscillation searches.  There
have been at least eight separate velocity studies, mostly single-site,
that have reported a hump of excess power around 0.5--1.5\,mHz.  However,
there is not yet agreement on the oscillation frequencies, although a
consensus is emerging that the large separation is about 55\,\muHz.

Considerable controversy has been generated by the non-detection of
oscillations in Procyon from photometry with the MOST satellite, reported
by \citet{MKG2004}.  Their interpretation has been strongly criticized by
\citet{BKB2005}, who maintained that the noise level in the MOST photometry
is too high for the signal to be detected.  Support for this is given by
space-based photometry with the WIRE satellite by \citet{BKB2005b}, who
reported a noise level lower than that of MOST\@.  By fitting to the power
density spectrum, they extracted parameters for the stellar granulation and
found evidence for an excess due to p-mode oscillations.  It is clear that
Procyon will remain an interesting (and controversial) target.

\begin{figure}[t!]
\resizebox{\hsize}{!}{\includegraphics{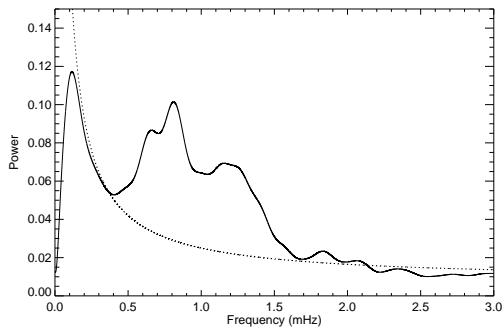}}
\caption{\footnotesize Smoothed power spectrum of Procyon, based on
  velocity measurements with the CORALIE spectrograph by \citet{ECB2004}.
  The dotted curve shows a fit to the background (instrumental, stellar
  granulation and photon noise) and the double-humped excess of power is
  due to p-mode oscillations.  }
\label{fig.procyon}
\end{figure}

Finally, we point out an interesting feature: all the published velocity
power spectra appear to show a dip at 1.0\,mHz (see \citealt{BMM2004} and
\citealt{CBL2005} for the most recent examples and \citealt{BKB2005} for a
full list of references).  This dip is visible in the raw plots but should
be even more pronounced if the power spectra are smoothed.  An example is
shown in Fig.~\ref{fig.procyon}, based on velocity measurements of Procyon
with the CORALIE spectrograph by \citet{ECB2004}, kindly provided by those
authors.  The dotted curve shows a fit to the background (instrumental,
stellar granulation and photon noise).  Above this, the excess of power due
to p-mode oscillations has a double-humped structure with a dip at 1\,mHz.
Interestingly, theoretical models by \citet{HBChD99} of a evolving star
with a similar mass and age to Procyon (their Fig.~4) show a dip in the
damping rate at a similar frequency.  If the dip in Procyon turns out to be
real, it opens the possibility of treating the shape of the oscillation
envelope as another observable that can be extracted from the power
spectrum and compared with theoretical models.

\section{\eboo}

This star, being the brightest G-type subgiant in the sky, remains a very
interesting target.  The claimed detection of oscillations almost decade
ago by \citet{KBV95}, based on fluctuations in Balmer-line
equivalent-widths, has now been confirmed by further equivalent-width and
velocity measurements by the same group \citep{KBB2003} and also by
independent velocity measurements with the CORALIE spectrograph
\citep{CEB2005}.  With the benefit of hindsight, we can now say that
\eboo{} was the first star for which the large separation and individual
frequencies were measured.  However, there is still disagreement on the
some of the individual frequencies, which reflects the subjective way in
which genuine oscillation modes must be chosen from noise peaks and
corrected for daily aliases.  Fortunately, the large separation is $\Dnu=
40$\,\muHz, which is half way between integral multiples of the
11.57-\muHz{} daily splitting (40/11.57 = 3.5).  Even so, daily aliases are
problematic, especially because some of the modes in \eboo{} appear to be
shifted by avoided crossings.

The first spaced-based observations of \eboo, made with the MOST satellite,
were presented at this meeting by J. Matthews and the paper has since
appeared on astro-ph \citep{GKR2005}.  The results have generated
considerable controversy.  \citet{GKR2005} showed an amplitude spectrum
(their Fig.~1) that rises towards low frequencies in a fashion that is
typical of noise from instrumental and stellar sources.  However, they
assessed the significance of individual peaks by their strength relative to
a fixed horizontal threshold, which naturally led them to assign high
significance to peaks at low frequency.  They did find a few peaks around
600\,\muHz{} that agreed with the ground-based data, but they also
identified eight of the many peaks at much lower frequency
(130--500\,\muHz), in the region of rising power, as being due to
low-overtone p-modes.  Those peaks do line up quite well with the regular
40\,\muHz{} spacing, but extreme caution is needed before these peaks are
accepted as genuine.  This is especially true given that the orbital
frequency of the spacecraft (164.3\,\muHz) is, by bad luck, close to four
times the large separation of \eboo{} (164.3/40 = 4.1).  A full discussion
of this issue is beyond the scope of this review, but it is safe to say
that the resolution to this controversy will probably have to wait for more
observations, both from the ground and from space.

\section{\muara}

This metal-rich G5~V star was already known to host a giant planet in a
2-year orbit \citep{BTM2001}, and was observed for oscillations using HARPS
by \citet{BBS2005}.  As a by-product, they found evidence for a second
planet of much lower mass, which they confirmed with follow-up observations
\citep{SBM2004}.  From the oscillation analysis, \citet{BBS2005} reported
43 modes with $l=0$--$3$, including evidence for rotational splitting of
$l=1$ modes.  If confirmed, this would be the first measurement of
rotational splitting in a star other than the Sun.

\citet{BVB2005} have compared the oscillations frequencies in \muara{} with
theoretical and discussed the possibility of deciding whether the high
metallicity in this star is primordial or due to pollution during planet
formation.  They concluded that it is not possible to decide this question
from the current data, although an accurate interferometric measurement of
the stellar radius could help.

\section{Other results}

The F9~V star \bvir{} has been observed in velocity by two groups:
\citet{MLA2004b} 
and \citet{CEDAl2005}.  Both found excess power from oscillations centred
at about 1.5\,mHz.  Only \citet{CEDAl2005} were able to measure the large
separation unambiguously, reporting a value of 72\,\muHz{} that is
consistent with one of the two possibilities suggested by \citet{MLA2004b}.
\citet{CEDAl2005} reported the detection of 31 individual modes, although
the single-site nature of these data means that, as always, some caution is
needed.  This is clearly a star worthy of further study.

The F5~V star HD 49933 (HR 2530; $V=5.7$), which is a target for the COROT
space mission, was observed over 10 nights with the HARPS spectrograph by
\citet{MBC2005}.  The star showed a surprisingly high level of velocity
variability on timescales of a few days.  This was also present as
line-profile variations and is therefore presumably due to stellar
activity.  The observations showed excess power from p-mode oscillations
and the authors determined the large separation ($\Dnu = 88.7\,\muHz$) but
were not able to extract individual frequencies.

Finally, we briefly mention the search for solar-like oscillations in red
giant stars with oscillation periods of 2--4 hours.  Recent ground-based
velocity observations by \citet{BdRM2004} with CORALIE and ELODIE
spectrographs have shown excess power and a possible large separation for
both $\epsilon$~Oph and $\eta$~Ser.  Meanwhile, earlier observations of
oscillations in $\xi$~Hya by \citet{FCA2002} have been further analysed by
\citet{SKB2004}, who found evidence that the mode lifetime is only about
2\,days.  If confirmed, this would significantly limit the the prospects
for asteroseismology on red giants.

\section{Conclusions}

We conclude with a few general points:
\begin{itemize}

\item Velocity observations are much less sensitive to the stellar
  granulation background than are intensity observations.  The main
  disadvantage with velocity is the need for sophisticated spectrographs.

\item Single-site data have severe limitations and multi-site observations
  are strongly preferred.

\item There are good reasons for trying to get better data for the
  brightest targets (\acena{} and~B, Procyon and \eboo), which are all
  interesting for different reasons.

\item Mode lifetimes have been estimated for a few stars.  It is very
  important to establish this parameter for a wide range of stars, from the
  main sequence though subgiants to the red giant branch.

\end{itemize}
Asteroseismology of solar-like stars has a bright future.  We must be sure
to choose targets carefully and arrange multi-site campaigns wherever
possible

\begin{acknowledgements}
We thank F. Carrier, P. Eggenberger and coworkers for providing the CORALIE
measurements of Procyon that were used to produce Fig.~\ref{fig.procyon}.
This work was supported financially by the Australian Research Council, by
the Danish Natural Science Research Council and by the Danish National
Research Foundation through its establishment of the Theoretical
Astrophysics Center.
\end{acknowledgements}

\end{document}